# El Niño–Southern Oscillation and Atlantic Multidecadal Oscillation Impact on Hurricanes North Atlantic Basin

Suchit Basineni

## Abstract

Tropical cyclones (TCs), including hurricanes and typhoons, cause significant property damage and result in fatalities, making it crucial to understand the factors driving extreme TCs. The El Niño–Southern Oscillation (ENSO) influences TC formation through tropospheric vorticity, wind-shear, and atmospheric circulations. Apart from atmospheric changes, oceans influence activity through sea surface temperatures (SSTs) and deep ocean heat content. These Atlantic SSTs determine the Atlantic Multi-decadal Oscillation (AMO), which indicates SST variability in the Atlantic. This research focuses on ENSO, AMO, and SST's impact on the strength and frequency of TCs in the North Atlantic Basin. AMO and SST anomalies are increasing at an alarming rate, but it remains unclear how their dynamics will influence future TC behavior. I used observational cyclone track data from 1950-2023, the Oceanic Niño Index (ONI), and NOAA's Extended Reconstructed SST V5 (ERSST). I found that Increasing SSTs over the past decade indicate stronger TCs, while warm-phase AMO periods correspond with higher TC frequency. Meanwhile, a greater frequency of landfalling TCs can be attributed to La Niña or ENSO-neutral, with El Niño decreasing the frequency of landfalling TCs. Such relationships suggest that as the seasonal predictability of ENSO and SSTs improve, seasonal TC forecasts may improve.

## 1  Introduction

Tropical Cyclones (TCs), or commonly known as hurricanes and typhoons, have caused significant damage through extreme winds, storm surge, and excessive rainfall, leading to widespread structural damage and flooding [1]. Each landfalling tropical cyclone in the United States has resulted in an average of $22.8 billion in property damage, caused business disruptions, and increased insurance claims [2]. The combination of economic loss, environmental destruction, and welfare endangerment make TCs one of the most, if not the most, impactful natural disasters.

Recent data suggests that TCs in the United States are becoming both more frequent and more intense. Five major hurricanes (Category 3+) made landfall in the United States from 2020-2023, while only three major hurricanes made landfall from 2010-2019. In the previous decade, 13 hurricanes (Category 1+) were recorded to make landfall, but 2020-2023 alone have yielded 10 hurricanes in the United States [3]. There is also evidence from the Accumulated Cyclone Energy (ACE) Index value, which incorporates TC strength and duration for each season, that in the past 30 years, a substantial uptick in TC intensity has been observed (Figure 1) [4].

It is therefore important to understand what factors may be influencing these changes in frequency and intensity. Three potential drivers of TC frequency and intensity in the North Atlantic basin are the El Niño–Southern Oscillation (ENSO), Sea Surface Temperatures (SSTs), and Atlantic Multi-decadal Oscillation (AMO) [5].

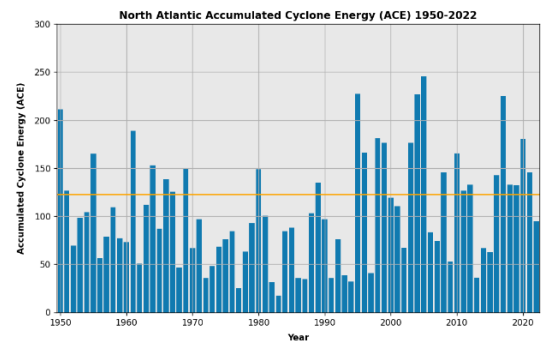

**Figure 1.** The Accumulated Cyclone Energy (ACE) index, which incorporates TC strength and duration for each season, suggests there has been an increase in North Atlantic TC intensity in recent decades. The Accumulated Cyclone Energy (ACE) is calculated annually by summing the squares of the estimated 6-hour maximum sustained tropical cyclone wind speeds (≥39 kt).The orange line indicates the average ACE value (123) based on the data from 1950-2022.

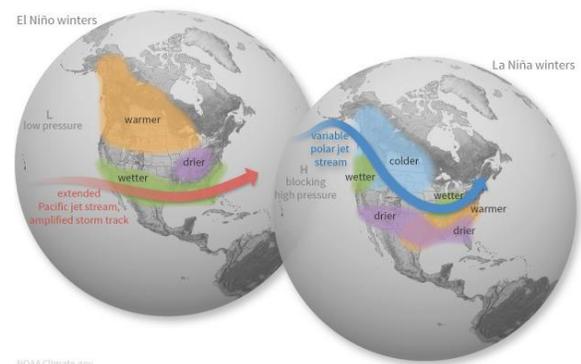

**Figure 2.** The El Niño–Southern Oscillation (ENSO) has had a considerable impact on North Atlantic TC activity. The figure [6] displays the typical climate patterns associated with El Niño and La Niña winters. ENSO typically has little impact in the summer in the US but has a significant influence in the Pacific and Atlantic Oceans [7].

The El Niño–Southern Oscillation (ENSO) (Figure 2) is a major factor influencing global tropical cyclone activity, indicating shifts in sea surface temperatures and atmospheric pressure across the Pacific Ocean [8]. ENSO has three phases: El Niño, La Niña, and Neutral. During El Niño, above-average SSTs are present in the Central and Eastern Pacific Ocean. Low-level winds over the equator may weaken or switch directions during El Niño. La Niña signals below-average SSTs in the Central and Eastern Pacific Ocean. The low-level winds moving east to west across the equator tend to increase. Neutral occurs when El Niño or La Niña is not present. In most cases, SSTs are near average but can be anomalous in certain cases. During stronger ENSO phases, both SSTs and atmospheric patterns must be conducive to that phase [9]. Even though ENSO is defined by the Pacific Ocean, it impacts the TC activity in the North Atlantic Basin. As ENSO impacts tropospheric vorticity, wind-shear, and atmospheric circulations, El Niño is associated with fewer TCs in the North Atlantic Basin. La Niña is associated with a greater amount of TCs in the North Atlantic Basin [10].

The Atlantic Multi-decadal Oscillation (AMO) (Figure 3) is another ocean oscillation index known to influence TC activity and reflects long-duration changes in SSTs in the North Atlantic [11]. The AMO has warm or cold phases that on average occur for 20-40 years.[5] Warm phases have been attributed to tropical cyclones maturing into stronger TCs than during cool phases [12]. In the past 1 ½ years, the AMO has been ~1.66 times greater than the previous peaks. This new data has led to a greater need for analysis of the impacts of these record-breaking SSTs without previous seasonal analogs to which to refer.

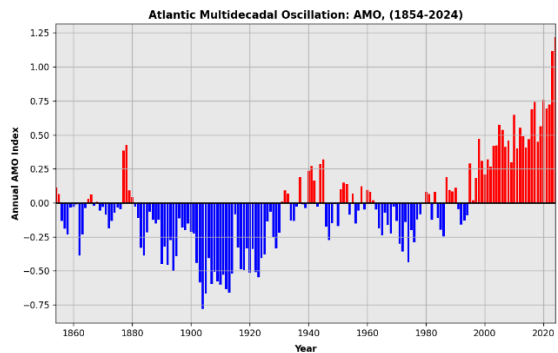

**Figure 3.** The Annual AMO Index is an important driver of Atlantic TC behavior. The figure displays the Annual AMO Index, which is calculated by subtracting the annual global mean SST excluding the North Atlantic Basin from the annual mean SST in the North Atlantic Basin. The data is displayed from 1854-2024.

Given the recent increases in cyclone intensity, an open research question is how much climate change has already played a role, and will impact TC activity in the future. According to the Intergovernmental Panel on Climate Change (IPCC), the increase of 1.1°C compared to the historical 1850-1900 average is attributed to an increase in peak wind speeds in TCs, more frequent and extremely heavy rainfall, and likely a higher rapid intensification rate in TCs [13]. Climate change has been linked to the rise in SSTs, but the ENSO and AMO patterns are considered to be naturally variable [14]. There is some evidence from climate models that climate change will lead to more intense El Nino and La Nina events but remains an active area of research [15]. However, the AMO is considered to be indicative of climate change occurring [16]. As these variables continue to shift significantly, research into the impact on TC formation is essential. This research focuses on ENSO, AMO, and SST's impact on the strength and frequency of TCs in the North Atlantic basin. I first identify trends in TC behavior since 1950 using observational data, then compare these trends against observed ENSO, AMO, and SST patterns. Lastly, I use a global climate model to examine future changes in SST patterns relative to historical conditions and assess what that may indicate for future TC behavior.

## 2  Methods

For this analysis, I used the publicly available NOAA's International Best Track Archive for Climate Stewardship (IBTrACS) dataset [17]. IBTrACS includes global tropical cyclone best-track data from various government reporting agencies. The dataset includes cyclone track information from 1851 to the present day, including the position coordinates and dates of tracks along with corresponding wind speeds and pressure provided at 3-hour increments globally. Data from 1950 to 2023 for the North Atlantic basin was utilized here, a total of 483 unique TC track records. Here I consider TCs to be any IBTrACS storm exceeding 64kt. The North Atlantic Basin spans from the Coastline of North America on the Eastern Pacific to 30°W. The dataset had a resolution of 0.1° or ~10 km.

For ACE data, I used CSU's monthly Atlantic Hurricane ACE values from 1851-2022 [18].

For AMO data, I used NOAA's AMO time series based on NOAA ERSSTV5 [19]. The dataset is the ERRST AMO Index in the North Atlantic from 0°N to 60°N.

For SST data, I used NOAA's Extended Reconstructed SST V5 (ERSST V5) monthly mean values from 1854 to 2023 [20]. The data is gridded on a 2°x2° (~222km) resolution (89x180), spanning 88°N to 88°S and 0°E to 358°E.

For ENSO analysis, I used NOAA's Oceanic Niño Index (ONI) dataset [21] The dataset represented a 3-month running mean of ERSST V5 SST anomalies in the Niño 3.4 region from 5° N to 5° S and 170° W to 120° W.

When looking at future SSTs, I utilized simulation data from the Coupled Model Intercomparison Project Phase 6 (CMIP6) [22]. CMIP6 includes 100 distinct climate models from 49 modeling groups. Using their future SST model data, I plotted the trends. The CMIP6 model used here to understand past, present, and future climate changes due to natural and human-induced variability is from the National Center for Atmospheric Research (NCAR) modeling center and provides monthly SST projections out to 2100 [23]. The projections are based on the High Emissions (ssp585) scenario, which assumes large increases in greenhouse gasses in the coming decades.

To conduct my analysis, I used the Python computing language to process and visualize the data. I used Python libraries such as pandas and xarray to analyze my data as well as matplotlib to plot the data. Data analysis included merging datasets or removing repetitive or missing information when comparing maximum wind speeds or related environmental factors on plots. Unless noted, all the plots displayed were created originally.

## 3 Results & Discussion

For the first part of this analysis, I investigated decadal North Atlantic tropical cyclone paths and the AMO during that decade. The goal of this is to identify trends in cyclone landfall locations, intensity, and frequency.

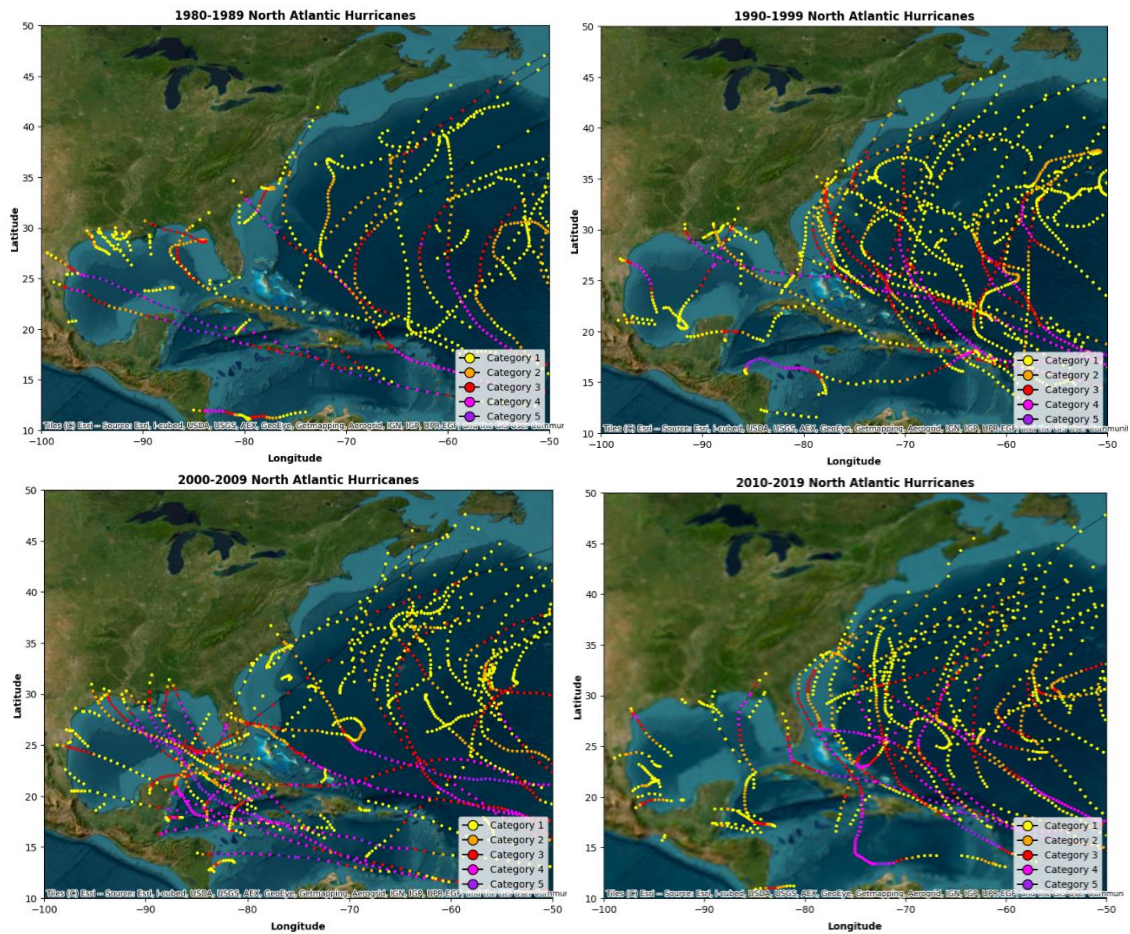

**Figure 4.** Observed cyclone track data suggests an increase in the frequency of Category 1-2 hurricanes during warm phase AMO years (1990-2019) compared to cold phase AMO years (1980-1989). The figure (10°N - 50°N, 100°W - 50°W) displays decadal North Atlantic Tropical Hurricane tracks from 1980-2019. Each hurricane is connected by a black line, indicating 3-hour intervals. Yellow

dots indicate Category 1 strength (64-82 kt), orange dots indicate Category 2 strength (83-95 kt), red dots indicate Category 3 strength (96-112 kt), pink dots indicate Category 4 strength (113-136 kt), purple dots indicate Category 5 strength (≥ 137 kt) [24].

As displayed in Figure 4, I investigated potential relationships between AMO and TC activity and compared the AMO time series with the corresponding TC tracks dataset.

### 3.1 AMO and TC Track Plot Takeaways

**Table 1.** The average amount of decadal hurricanes is increased during warm phases. The figure is a summative data table including information from figures 3 and 4.

|  | Gulf Of Mexico Landfalls | | Atlantic Coast Landfalls | | | |
| --- | --- | --- | --- | --- | --- | --- |
|  | Hurricanes | Major Hurricanes | Hurricanes | Major Hurricanes | Decadal ACE | AMO Phase |
| 1950-1959 | 14 | 3 | 12 | 3 | 1096.3 | Warm |
| 1960-1969 | 10 | 6 | 6 | 0 | 1124.1 | Cold |
| 1970-1979 | 10 | 7 | 4 | 0 | 656.9 | Cold |
| 1980-1989 | 15 | 4 | 5 | 1 | 778.3 | Cold |
| 1990-1999 | 10 | 3 | 8 | 1 | 1071.4 | Warm |
| 2000-2009 | 19 | 7 | 10 | 1 | 1300.7 | Warm |
| 2010-2019 | 11 | 4 | 6 | 0 | 1222.4 | Warm |
| 2020-2023 | 11 | 5 | 2 | 0 | 568.6 | Warm |
| **Average Decadal Hurricanes** (Gulf of Mexico + Atlantic) | | | | | | |
| Cold Phase | 16.677 | | | | | |
| Warm Phase | 22.5 | | | | | |
| **Average Decadal Major Hurricanes** (Gulf of Mexico + Atlantic) | | | | | | |
| Cold Phase | 6 | | | | | |
| Warm Phase | 5.5 | | | | | |

Note: 2020-2023 is not included in the total mean value. Major Hurricanes are Cat 3+. Only accounts for contiguous U.S. Landfalls.

TC landfalls were split into two categories: Gulf of Mexico landfalls and Atlantic Coast landfalls. During cold phases and warm phases, more Gulf TCs made landfall each decade. Additionally, there were significantly more landfalling major hurricanes in the Gulf of Mexico compared to the Atlantic. While this trend may seem set in stone regardless of the AMO, the numbers will likely increase in the Atlantic and Gulf as SSTs continue to increase.

Taking the average decadal hurricanes from the Gulf of Mexico and Atlantic combined, there are approximately 17 hurricanes during cold phase decades whereas warm phase decades averaged around 22.5 hurricanes. As we continue to experience a prolonged and stronger warm phase AMO (Figure 3), the results here suggest that average decadal landfalling hurricanes are likely to increase too. This trend is consistent with recent data from 2020-2023 which has had 13 landfalling hurricanes.

On the other hand, when looking at the average decadal major hurricanes, cold phases average 0.5 more major hurricanes per decade. Even though this data may be contradicting, 2020-2023 has had 5 major hurricanes and record AMO anomalies, indicating this recent spike may promote a greater number of major hurricanes.

Another indicator of TC frequency, decadal ACE has indicated significantly higher activity in warm phase periods compared to cold phase periods (Table 1). Except for 1950-1969, ACE values were ~400 higher.

2000-2009 had almost double the ACE as 1970-1979 (Table 1). ACE from 2020-2023 has already almost surpassed cold phase years in four years. Overall this indicates an increase in ACE in recent decades, with elevated ACE during AMO warm phase periods.

Higher ACE values and mean decadal hurricane values suggest that warm-phase AMO may have more of a significant impact on the frequency and intensity of TCs, specifically category 1-2 hurricanes. Some limitations may include only taking North American continental landfalls and excluding post-tropical cyclones (PTCs) in the analysis. Regardless of these limitations, the AMO has nearly had a 66% (+0.5) increase in anomalies in the past two years. More data will need to come in future years to confirm these large upticks in recent years.

### 3.2 ENSO and Tropical Cyclones

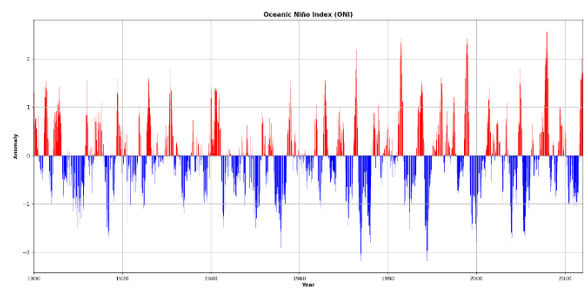

**Figure 5.** The Oceanic Niño index has had more extreme anomalies in the past 50 years. The figure displays the Oceanic Niño Index (ONI) from 1900 to 2023. The Oceanic Niño Index is one key indicator of the current ENSO pattern. Anomalies ≥0.5 indicate El Niño; -0.5 to 0.5 indicates neutral; anomalies ≤0.5 indicate La Niña.

As mentioned in the introduction, ENSO impacts SSTs in the Pacific as well as the atmospheric patterns across the Pacific and in the North Atlantic. The Oceanic Niño Index (ONI) is the main factor in determining the ENSO phase. The ONI is calculated using a 3-month rolling average for SSTs over the Niño 3.4 region. The anomaly had been traditionally calculated using the previous 30-year SST average over the Niño 3.4 region [25]. In recent years, the index has adapted to the rate of rising SSTs by taking the 30-year average centered around the first year of a 5-year period. This means that current data is preliminary till 10 years of future data is present [26]. Niño 3.4 region is located in the center of the Pacific along the equator [25]. As depicted by Figure 5, extreme variations have become more common after 1970. Anomalies have dipped to -2.18 in 1973 and 1988 indicating a very strong La Niña. Anomalies have soared to 2.57 in 2017 indicating an extremely strong El Niño. As these variations have become greater in recent decades, it indicates these patterns will have a greater impact on hurricane season in the North Atlantic Basin. Another trend that is denoted from 1990 onwards is that certain El Niño seasons have only become more anomalous compared to La Niña seasons. As ENSO is an index that adjusts per the rise of SSTs, La Niña anomalies have seemingly decreased. Still, the strongest La Niña's (indicating the coldest temperatures) have been warmer overall than the strongest El Niño's in previous decades.

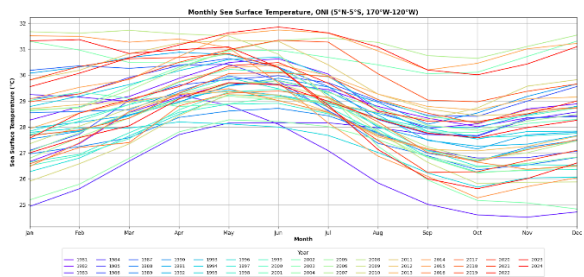

**Figure 6.** Compared to average SSTs, considerably larger SST anomalies have occurred after 2000. The figure displays the Monthly Sea Surface Temperatures (°C) in the ONI region (5°N-5°S, 170°W-120°W). Cooler colors (purple and blue) represent 1981-1999. Shades of green represent 2000-2010. Warmer colors (yellow, orange, and red) represent 2011-2024.

These warmer SSTs are evident in the region where the ONI is calculated. After 2000, certain years have had significantly warmer SSTs from July to February indicating larger spikes in ONI anomalies (Figure 6). As discussed before, the rate at which global SSTs are increasing is far outpacing the previous 30-year running averages, which is causing lesser extremes during La Niña.

Even though these lesser extremes during La Niña are occurring, it does not necessarily indicate the lower frequency of TCs. Understanding the implications of rising SSTs on ENSO is important in considering all factors conducive to a greater frequency and intensity of hurricanes in the North Atlantic Basin.

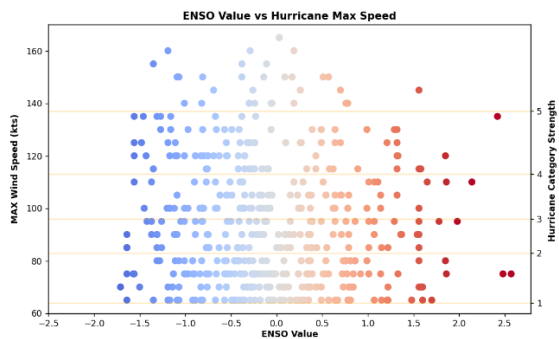

**Figure 7.** La Niña and Neutral phases are conducive for stronger max-speed hurricanes. The figure displays the maximum wind speed in knots for TCs that achieved Category 1 or greater strength.

As La Niña supports favorable conditions in the North Atlantic Basin, there have been more hurricanes than El Niño from 1950-2023 (Table 2). ENSO-neutral conditions have yielded the largest amount of hurricanes (Table 2). Regardless of hurricane count, La Niña and El Niño seasons have produced storms with a higher ceiling for peak intensity. La Niña's ceiling is 165 kt, the Neutral phase' ceiling is 170 kt, and El Niño's ceiling is 155 kt (Figure 7). Additionally, during extremely strong El Niño phases, there is a large decrease in the average max windspeed for unique hurricanes.

### 3.3 Frequency of Hurricanes during respective ENSO Phase

**Table 2.** El Niño has the highest number of hurricanes/year in the North Atlantic Basin. The number of hurricanes does not include weakening and restrengthening hurricanes. It only includes unique entries.

|  | La Niña | Neutral | El Niño |
|---|---|---|---|
| **Total Length** (years) | 22.583 | 34.583 | 16.833 |
| **# Of Hurricanes** | 150 | 215 | 117 |
| **Hurricanes/Year** | 6.642 | 6.217 | 6.950 |

The Neutral ENSO phase is the dominant phase as SSTs vary frequently in the ONI region. The neutral phase lasted 34.583 years with 215 hurricanes 1950-2023. La Niña lasted 22.583 years with 150 hurricanes and El Niño lasted 16.8333 years with 117 hurricanes. Contrary to expectations, El Niño had the highest value at 6.950 hurricanes per year in the North Atlantic Basin with La Niña trailing at 6.642 and the neutral phase at 6.217. Even though this may indicate El Niño has produced the greatest mean amount of hurricanes, North American landfalling hurricanes are the least during this phase.

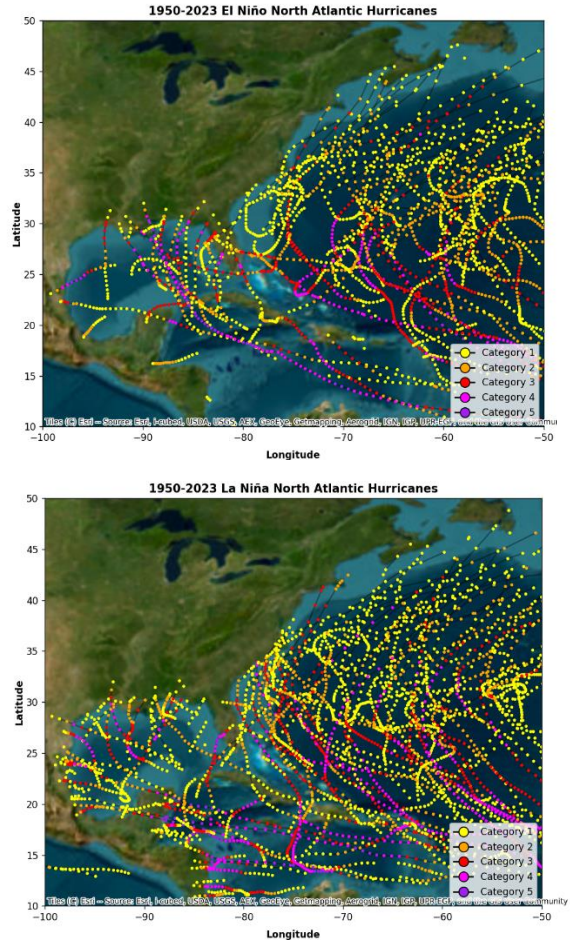

**Figure 8.** There are a greater number of landfalling hurricanes during La Niña compared to El Niño. This figure displays North Atlantic hurricanes from 1950-2023 during La Niña or El Niño.

Hurricane landfall probabilities are significantly increased in western parts of the Gulf of Mexico, Central America, and the East Coast (Figure 8). This supports that La Niña is conducive to more landfalls on the North American Continent and surrounding islands, but it does not increase overall TC activity across the whole North Atlantic Basin. As this continues, rising SSTs will only increase the risk of these landfalling hurricanes during La Niña seasons.

### 3.4 SSTs and Tropical Cyclones

SSTs are another major factor of TC intensity in the North Atlantic. As ENSO primarily deals with the frequency of hurricanes in certain areas, SSTs increase the risk of stronger hurricanes [27].

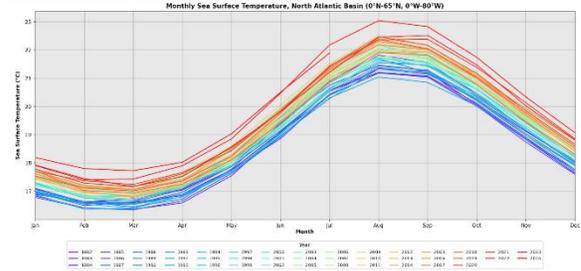

**Figure 9.** The SSTs in the North Atlantic have been increasing at higher rates than previous years. The monthly sea surface temperatures in the North Atlantic Basin (0°N-65°N, 0°W-80°W) from 1982-2024.

The North Atlantic SSTs are increasing at higher rates than in previous years (Figure 9). These new trends are a cause for concern in the Atlantic. SSTs are generally required to be greater than 26 degrees to sustain and support hurricanes [28]. These warmer temperatures can increase the severity and rate of intensification in these storms (Figure 11). The global increase in SSTs does not correlate to a greater number of hurricanes though, owing to the complexities in the relationship between global SSTs and TC activity.

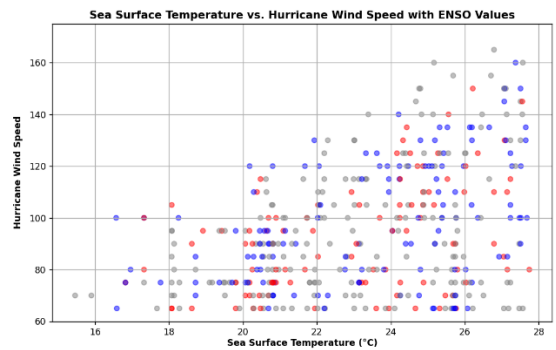

**Figure 10.** As SSTs increase, the maximum strength of a hurricane is significantly higher. This used the 2°x 2° Extended Reconstructed SST V5 (ERSST V5) monthly mean values. These are hurricanes from 1950-2023). Localized SST values may be higher than what is depicted, but overall SST values over a 222 km x 222 km area are used for each data point.

ENSO variability has a large influence on the frequency and areas of hurricane development. As noted before, La Niña and Neutral Phase hurricanes tend to have greater maximum strength compared to El Niño hurricanes (Figure 10). Over the entire North Atlantic Basin, the frequency of El Niño and La Niña hurricanes. tends to be similar. This raises the question of how strong future hurricanes may become.

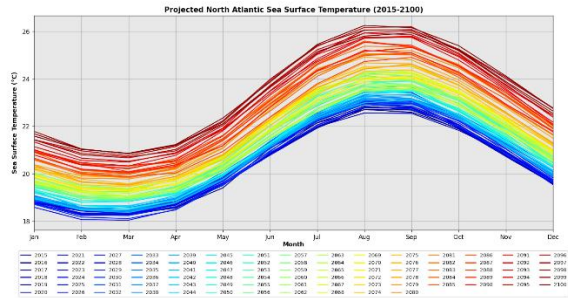

**Figure 11.** The rate at which the North Atlantic Basin's (0°N-65°N, 0°W-80°W) SSTs are rising is expected to be higher than previous years. This is the CMIP6 Model high emissions scenario.

Peak SSTs in a scenario up until 2100, depict North Atlantic SSTs reaching 26°C near 2100 (Figure 11). Even though this is a high-emission scenario, 2023 SSTs could be compared to the model's depictions of North Atlantic SSTs around 2040. This extremely alarming trend is an area that will continue to pose more questions in hurricane intensity, as well as require the use of new model data to depict worse-case scenarios.

## 4  Conclusion

The AMO, SSTs, and ENSO patterns play a large role in hurricane development yearly. These patterns help determine the environment in the North Atlantic Basin. TC behavior since 1950 in the North Atlantic Basin has yielded more TCs in the Gulf than in the Atlantic, but the total amount of TCs per decade is likely to rise. The study that warm-phase AMO periods may significantly impact TC intensity and frequency, La Niña conditions promote more landfalling hurricanes, and TC frequency suggests that SSTs impact intensity, not frequency. SSTs are increasing at a rate higher than modeled by various scenarios, prompting further research on the impact of more extreme SST anomalies. Additionally, global warming has had an impact on the AMO, indicating record-breaking anomalies in the past 1 ½ years, as well as on SSTs. These anomalies continue to support a greater amount of intensification, leading to a higher frequency of category 1-2 hurricanes landfalling as well as a great frequency of storms during warm phases.

This analysis does not include landfalling Post-Tropical Cyclones at TC strength, which may impact the overall TC activity per decade. Additionally, data from 2023-2024 presented may be reanalyzed, which may slightly alter values in future charts. TC SSTs were taken over larger areas due to processing limitations, which may slightly impact the accuracy.

Significant increases in SSTs and climate variability will continue shortly, requiring constant updates and research on new trends. This study highlights the need to continue the analysis of ENSO and AMO impacts on TCs in the North Atlantic. Future work will aim to incorporate more time series analysis and new data to determine the magnitude of impact on TCs.

## 5  Acknowledgements

I would like to acknowledge and thank Dr. Tristan Ballard for guiding me in my research with his expertise in the field.

# 7 Authors

Suchit Basineni is a 12th grader at High Technology High School in New Jersey. He has a strong passion for atmospheric science and climatology. He hopes to be a data scientist in the field.